\documentclass[12pt,preprint]{aastex}

\usepackage{subfigure}
\usepackage{longtable}
\usepackage{graphicx}
\usepackage{graphics}
\usepackage{keyval}
\usepackage{trig}

\begin{document}
\title{Self-consistent models of triaxial galaxies in MOND gravity}
\author{You-gang Wang$^{1,3}$, Xufen Wu$^{2}$, HongSheng Zhao$^{1,2}$}
\altaffiltext{1}{National Astronomical Observatories, Chinese
Academy of Sciences, Beijing 100012 } \altaffiltext{2}{ SUPA, School
of Physics and Astronomy, University of St Andrews, KY16 9SS, UK }
\altaffiltext{3}{Department of Astronomy Peking University, Beijing
100871, China} \email{hz4@st-andrews.ac.uk}

\begin{abstract}
The Bekenstein-Milgrom gravity theory with a modified Poisson
equation is tested here for the existence of triaxial equilibrium
solutions. Using the non-negative least square method, we show that
self-consistent triaxial galaxies exist for baryonic models with a
mild density cusp $\rho \sim {\Sigma \over r}$. Self-consistency is
achieved for a wide range of central concentrations, $\Sigma \sim
10-1000\mathrm{M_{\odot}pc^{-2}}$, representing low-to-high surface
brightness galaxies. Our results demonstrate for the first time that
the orbit superposition technique is fruitful for constructing
galaxy models beyond Newtonian gravity, and triaxial cuspy galaxies
might exist without the help of Cold dark Matter.
\end{abstract}


\keywords{galaxies: kinematics and dynamics- methods: numerical - dark matter}

\section{Introduction}

Constructing models of galaxies in triaxial equilibrium is a
classical challenge in dynamics, either in Newtonian or Modified gravity.
However, due to numerical
constructions, few such models have been developed in Newtonian since the
pioneering work of Schwarzchild(1979) on triaxial elliptical galaxies, and
Zhao (1996) on the fast rotating triaxial bar of the Milky Way.
Merritt \& Fridman (1996) (hereafter
MF96) considered a model with a central density cusp $r^{-1}$ and
found that it is possible to construct a model of triaxial equilibrium galaxies self-consistently in
Newtonian dynamics.  Models with a steeper density cusp are not in equilibrium.
This has generated interests in understanding cusp-triaxiality
relation, the triaxiality-velocity anisotropy relation and the
cusp-black hole relation. Recently, Capuzzo-Dolcetta et al. (2007)
modeled $r^{-1}$ cuspy triaxial galaxies in $\Lambda$ Cold Dark Matter (CDM) haloes. They took the model of MF96 as the luminous density
distribution, and found that a model of cupsy triaxial galaxies with CDM
halos is also self-consistent.

A tough problem for $\Lambda$CDM ($\Lambda$ plus Cold Dark Matter)
is that there is no physics to justify $\Lambda$ the tiny
cosmological constant or dark energy (see White 2007, Sarkar 2007).
Although future particle physics experiments might well prove the
existence of new species of particles and vacuum energy, their
experimental-determined abundance could easily be factors of a few
different from the 1:3 ratio as precisely required by fitting
observations of Microwave background (Zhao 2006).  This motivates
exploring alternative theories like the co-variant version of
Modified Newtonian Dynamics (MOND) in the mean-time.  E.g., instead
of more traditional interpretation of modified gravity (Bekenstein
2004), MOND was given the interpretation as a co-variant Dark Energy
(DE) the ${\bf V\Lambda}$ model of Zhao (2007).  This model uses
General Relativity plus a non-uniform Dark Energy (DE) fluid
(described by a four-vector flow) to give both the effects of DE and
DM.  Perturbations of such DE fluid bends space-time as an effective
DM without actually invoking DM.  Unlike $\Lambda$CDM halos, which
are positive and nearly round, the effective DM can sometimes be
extremely flattened or negative in some regions of galaxies or
clusters (Wu et al. 2007, 2008), although the data are not good
enough to tell such negative regions yet (Nipoti et al. 2007a).

Overall MOND and $\Lambda$CDM halos are comparably successful in
explaining the flat rotation curves of high surface brightness discs
and satellites orbits at large radii (e.g., Angus et al. 2008).
However, both theories are faced with severe challenges:
$\Lambda$CDM hinges on uncertain feedback on galactic scales, and
even the maximum feedback falls short of explaining the velocity
curves of low-surface brightness systems (Gnedin \& Zhao 2002);
Bekenstein's (2004) co-variant MOND seems to rely on
non-relativistic neutrinos on large scale to match the weak lensing
data (Angus et al. 2007) and the Cosmic Background Radiation
(Skordis et al. 2006). \footnote{although these problems appear
resolvable in the ${\bf V\Lambda}$ model (Zhao 2007).}

The tight correlation observed between the mass profiles of baryonic
matter and dark matter at all radii in spiral galaxies (e.g.,
McGaugh et al. 2007; Famaey et al. 2007) is still the best case for
the MOND paradigm of Milgrom (1983), which postulates that for
accelerations below $a_0 = 1.2 \times 10^{-10} {\rm m} \, {\rm
s}^{-2}$ the effective gravitational attraction approaches $(g_N
a_0)^{1/2}$, where $g_N$ is the usual Newtonian gravitational field.
Indeed, without resorting to galactic dark matter, this simple
prescription is amazingly successful at reproducing galactic
rotation curves over five decades in mass ranging from tiny dwarfs
(e.g., Gentile et al. 2007) to early-type disk galaxies (e.g.,
Sanders \& Noordermeer 2007) to massive ellipticals (Milgrom \&
Sanders 2003). Some outliers exist in models of gravitational
lensing (Zhao et al. 2006, Chen \& Zhao 2006, Shan et al. 2008).
Lensing and structure formation are well-defined calculations in
co-variant MOND (Halle \& Zhao 2007). Galaxy formation simulations
in classical MOND have been carried out too, and it is possible to
form bars (Tiret \& Combes 2007), and elliptical galaxies by mergers
(Nipoti et al. 2007b), and spherical bulges by instability (Zhao, Xu
\& Dobbs 2008).

Very little is known about the triaxial stationary equilibrium in non-Newtonian dynamics,
although one speculates that such equilibrium might exist.
In this paper, we test the existence of triaxial galaxies in the Bekenstein-Milgrom
MOND theory, a well-defined classical theory with
relativisitic and cosmological extentions.
We apply the same density model given by MF96 with a fixed mass, and axis ratios.
Our aim is to examine whether this model is self-consistent in MOND.

The MOND theory is scale-dependent.  When the gravitational acceleration is below $a_0$, the scaling
deviates from the $r^{-2}$ Newtonian law.  Here we consider models with different ratio of
$a_0$ to the simulation unit G $\times$ (Total Mass) / (Scale Radius)$^{2}$.
We start with a nearly Newtonian system (high surface brightness)
and finish with a deep-MOND system (low surface brightness).

The rest of the paper is organized as follows. In \S2, we present
the density distribution of the galaxy and the corresponding
potential and forces. The orbits are calculated in \S3. The
construction of self-consistent models is discussed in \S4, and in \S5,
we give our conclusions and disscussions.

\section{The density, potential distribution and forces in the systems of triaxial galaxies}
\begin{figure}
\plotone{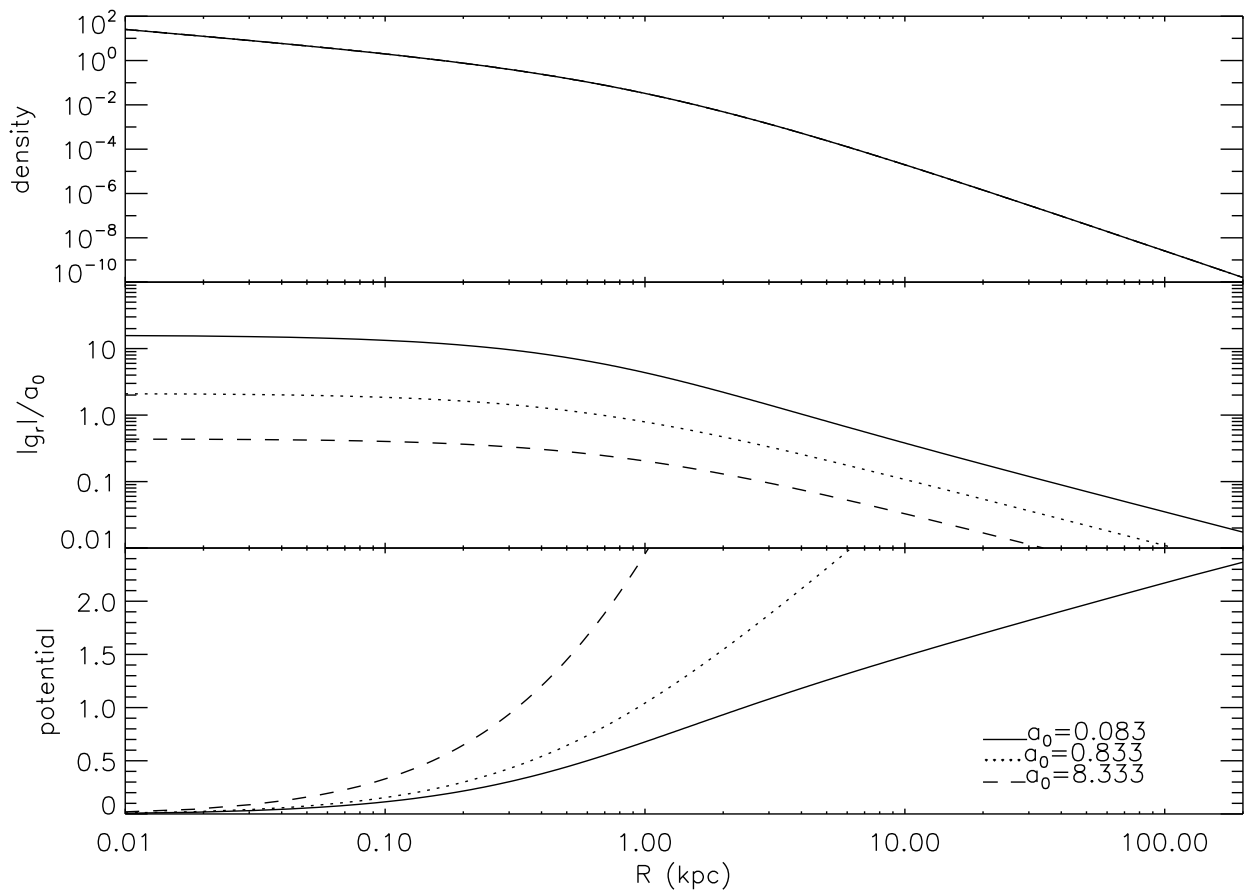} \caption{Upper: the radial distribution of the
dimensionless density $\rho$ in units of $M/(2\pi h_xh_yh_z)$, which
is identical for the three models. Middle: the MOND gravity strength
$|g_r|/a_0$ as a function of $R$ along the long axis. Lower: the
potential major axis profile, for different values of $a_0$,
obtained from the Poisson solver.} \label{den}
\end{figure}

The baryon component adopted by Capuzzo-Dolcetta (2007) has the
following ellipsoidal mass distribution:
\begin{equation}
\rho_b(R)=\frac{M}{2\pi h_x h_y h_z}\frac{1}{R(1+R)^3},
\end{equation}
where
$R^2=\frac{x^2}{h_x^2}+\frac{y^2}{h_y^2}+\frac{z^2}{h_z^2}(h_z\leq
h_y\leq h_x)$ is the length of the major axis, and $M$ is the total
baryon mass. For the ratio of the three major axes, we adopt
$h_x:h_y:h_z=1:0.86:0.7$ throughout this paper. This density profile
was studied in MF96, but in their work, it represented the density
profile of the dark matter. The density profile is plotted in the
top panel of Fig.~\ref{den}. Density is cuspy and scales as $R^{-1}$
at the center and decreases quickly as $R^{-4}$ at large radii.

Hereafter, we adopt units in which the total mass $M$, the x-axis
scale length $h_x$, and the gravitational constant $G$ are unity.
Thus one unit of time corresponds to
\begin{eqnarray}
T&&=G^{-1/2}h_x^{3/2}M^{-1/2}\nonumber \\
&&=1.49\times10^6\mathrm{yr}\bigg(\frac{M}{10^{11}M_{\odot}}\bigg)^{-1/2}\bigg(\frac{h_x}{\mathrm{1kpc}}\bigg)^{3/2}
\end{eqnarray}
The MOND nonrelativistic field equation is different from
Poisson's equation in Newton's theory (Bekenstein \& Milgrom 1984),
and it has the following form
\begin{eqnarray}\label{ps1}
4\pi G \rho&=&\nabla\cdot\left(\mu\nabla\Phi\right)\nonumber\\
           &=&\partial_x\left(\mu\partial_x\Phi\right)+\partial_y\left(\mu\partial_y\Phi\right)+\partial_z\left(\mu\partial_z\Phi\right),
\end{eqnarray}
where $\Phi$ is the gravitational potential generated by the density
distribution $\rho$. We adopt the function $\mu$ as in Zhao \& Famaey (2006),
which is given by
\begin{equation}\label{ps2}
\mu(g)=\frac{g}{g+a_0},
\end{equation}
Clearly $\mu=1$ for $g\to\infty$ and $\mu=g/a_0$ for $g\to0$, as
expected for any MOND theory.  The gravity strength $g$ is given by
\begin{eqnarray}\label{ps3}
g&=&{\left|\nabla\Phi\right|}\nonumber\\
 &=&\sqrt{(\partial_x\Phi)^2+(\partial_y\Phi)^2+(\partial_z\Phi)^2}.
\end{eqnarray}

From equation~(\ref{ps3}) it can be seen that the
equation~(\ref{ps1}) does not reduce to Poisson's equation if $a_0$
is not small enough. We solve equation~(\ref{ps1}) numerically,
using the Bologna group's MOND Poisson Solver (Ciotti et al. 2006),
which is based on a spherical grid. This code yields results which
are consistent with that of the Cartesian grid code of the Paris
group (Tiret \& Combes 2007). We apply the Poisson Solver with a
resolution of $256\times64\times128$ grids points, and choose the
radial grid points as
$r_i=2.0\mathrm{tan}[(i+0.5)\frac{0.5\pi}{256+1}]\mathrm{kpc}$. We
obtain the components of gravity in x, y, z directions, ie. values
of $\partial_x$, $\partial_y$, and $\partial_z$ from the Poisson
Solver.

In the lower panel of Figure 1, we show the potential radial profile
along the major axis for different values of $a_0$ in simulation
units. In the middle panel we show the radial gravity $|g_r|/a_0$ as
a function of $R$ along the long axis. In the simulations $a_0$ is
dimensionless and is equal to $0.0833$, $0.833$ and $8.33$, which
correspond to simulation units of $m=1.0\times 10^{10}M_{\odot}$,
$1.0\times 10^{9}M_{\odot}$, and $1.0\times 10^{8}M_{\odot}$
respectively, and $G=1$.
In all simulations, we start with a total mass of $M=10m$ (See Table
\ref{parameter}). We find that there is no obvious difference
between the mild and weak gravity cases, except in the central
region. For the strong gravity case ($a_0=0.083$), the gravitational
potential falls quickly at the outer regions of the galaxy.

\begin{figure}
\resizebox{16cm}{!}{\includegraphics{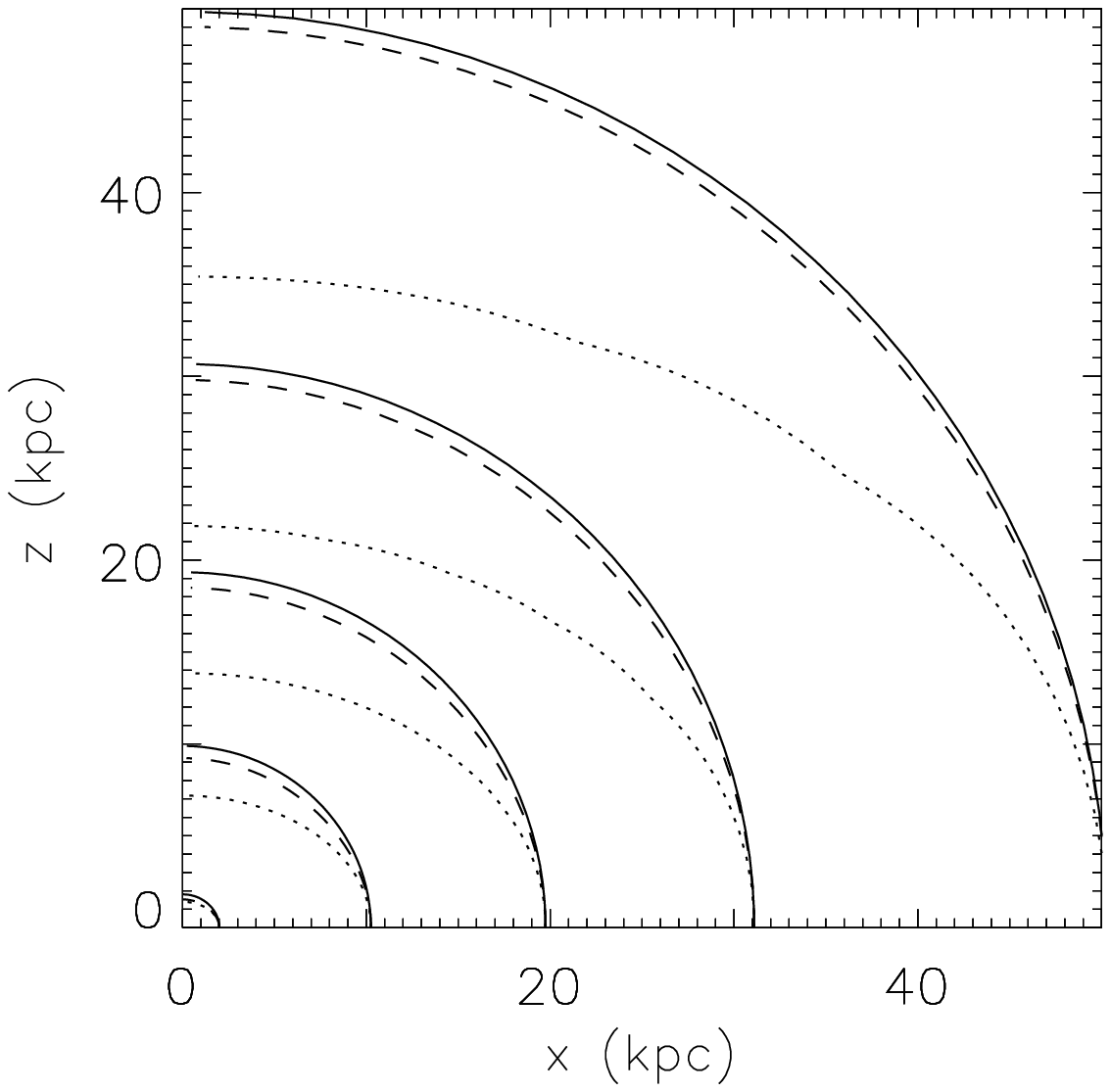},\includegraphics{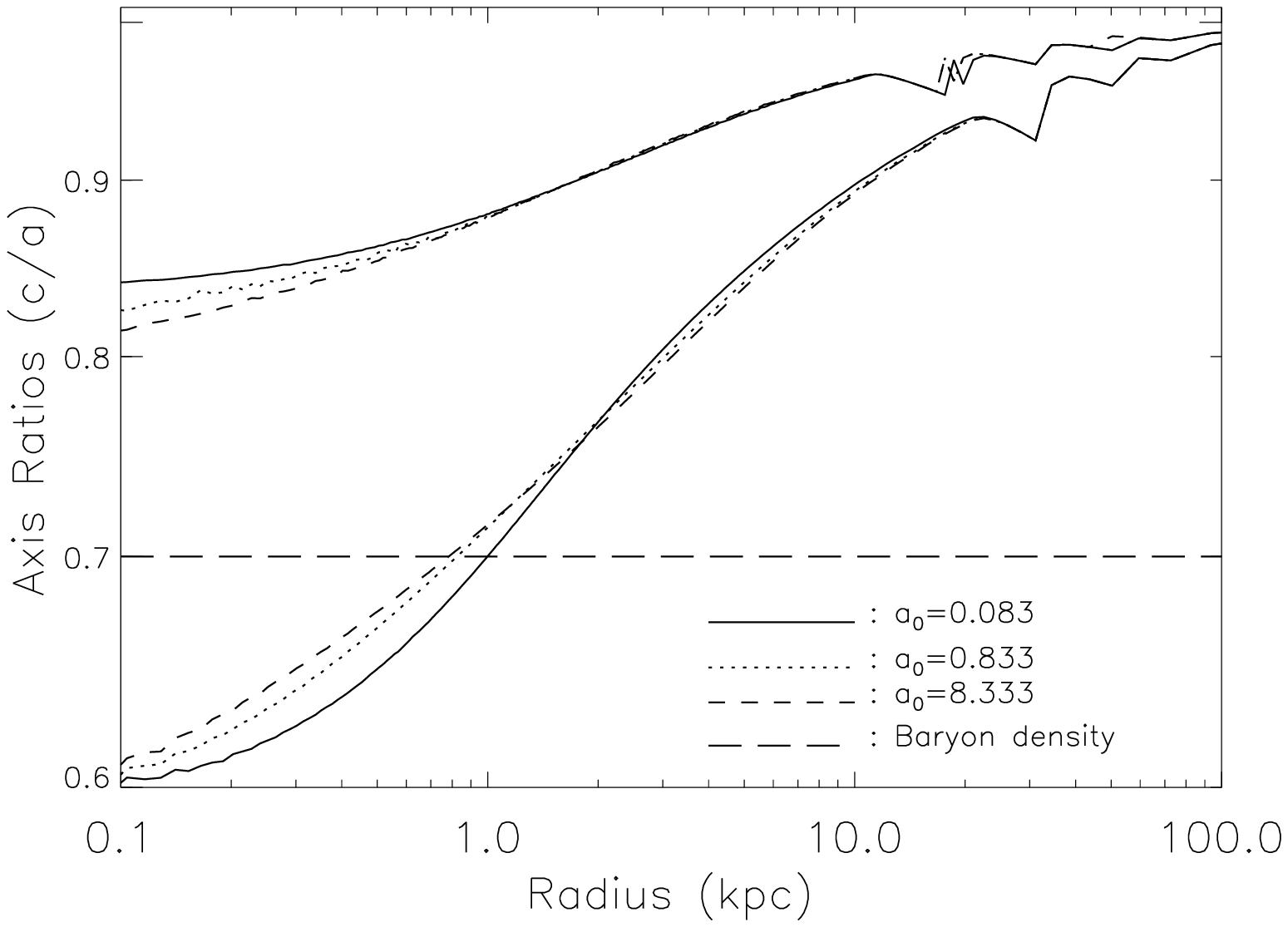}}\vskip
0.5cm \caption{Left panels: Isopotentials (solid lines),
isodensities of effective dark matter(dashed) and isodensities of
baryons (dotted) on the x-z plane, in the model of $a_0=0.833$. The
contours of the other two models are almost same. The isopotentials
are essentially a snapshot of the metric perturbations, and the
isodensities of effective DM halos are images of the perturbation of
certain Dark Energy fluid (Zhao 2007). Right panel shows the axis
ratio $c/a$ (the ratio between the short and long axis) of the
potentials (upper series of curves), effective DM densities  (lower
series of curves) and baryon densities(the long-dashed line) in a
log-log space. For each series, the solid, dotted and dashed lines
are for different models of $a_0$ as in the legend. The oscillations
at large radii are due to the numerical resolution. } \label{iso}
\end{figure}

The left panel of Figure \ref{iso} shows the isopotentials (solid
lines) in the gas with axial ratios $h_x:h_y:h_z=1:0.86:0.7$ at five
different radii, $R=$2, 10, 20, 30 and 50 kpc. There are no obvious
differences with different values of $a_0$(see Fig.~\ref{iso}b).
Further more, comparing with the density contours (the dotted lines
on the Fig.~\ref{iso}a), the isopotential contours are more
spherical than the isodensity (Fig.~\ref{iso}b), which is expected
(see also the examples in Lee \& Suto (2003) and Wang \& Fan
(2004)). The three-dimensional potential distribution is
approximately ellipsoidal, with eccentricity increasing from the
center to the outer part of the cluster. For example, the axial
ratio of potential distribution is $h_z^p:h_x^p=0.90:1.0$ at 2kpc,
while $h_z^p:h_x^p=0.99:1.0$ at 100~kpc. The dashed lines are
isodensities of the effective DM halo in MOND. On the
Fig.~\ref{iso}b, it is interesting that the axis ratio of effective
DM densities are smaller than that of baryon densities at radii
smaller than 1~kpc, where baryons dominate the dynamics. To produce
the same dynamical behavior(e.g. the rotation curves of galaxies),
the CDM halos always adopt a constant axis ratio of dark matter
density in the inner and outer part of galaxies, otherwise the CDM
models would be much too complex.

For the model of the elliptical galaxy, one important parameter is
the effective radii. Hernquist (1990) shew that the effective of the
elliptical galaxy is $\sim1.8153h_x$ for a spherical system. In our
models, the effective radii is also $\sim 1.8h_x$ because this
parameter is independent of gravity and is set by the baryon density
profile.

MOND is equivalent to an effective DM theory.  The effective density can be
obtained from
\begin{equation}
4\pi G \rho_{DM}^{eff} = \nabla \cdot \left[ (1-\mu)  \nabla \Phi
\right].
\end{equation}
This density can be quite different from both the baryonic density
and also the nearly ellipsoidal density of
the CDM halo.  A example is shown in Fig. \ref{iso}.

\section{Integration of orbits}

In Newton's theory, energy, and the three components of the angular
momentum, are integrals of motion in a spherical system. However,
only energy is kept as a constant in the triaxial potential. MOND
does not change the number of constants of motion.
Bekenstein-Milgrom's theory conserves energy for a stationary
potential. In order to obtain a specified accuracy, we use the 7/8
order Runge-Kutta algorithm (Fehlberg 1968) to carry out the
integrations. The integration time for a full orbit is 100 dynamical
times. One dynamical time is defined as the period of the 1:1
resonant orbit in the x-y plane. Table~\ref{shell} lists the periods
of 1:1 resonant orbits in the x-y plane as a function of energy.

In order to produce a library of all possible orbits, we followed Schwarzschild (1993) and MF96 in
selecting our initial conditions. Specifically, the definition of
start-space is the same as in MF96. The orbital catalogues include both
initially stationary orbits and x-z plane launched orbits with
nonzero velocity $v_y$, where $v_y=\sqrt{2(E-\Phi)}\neq0$ is the
velocity along the y-axis. For each model we have 20 energy surfaces.
The surfaces are the critical surface of the shells which divide the
total system into 21 parts of equal mass. One energy consists 342
initial conditions, where 192 orbits are from the stationary start-space
and the rest are from the x-z start-space. Therefore, for each model we
have 6840 orbits.

\begin{figure}
\plotone{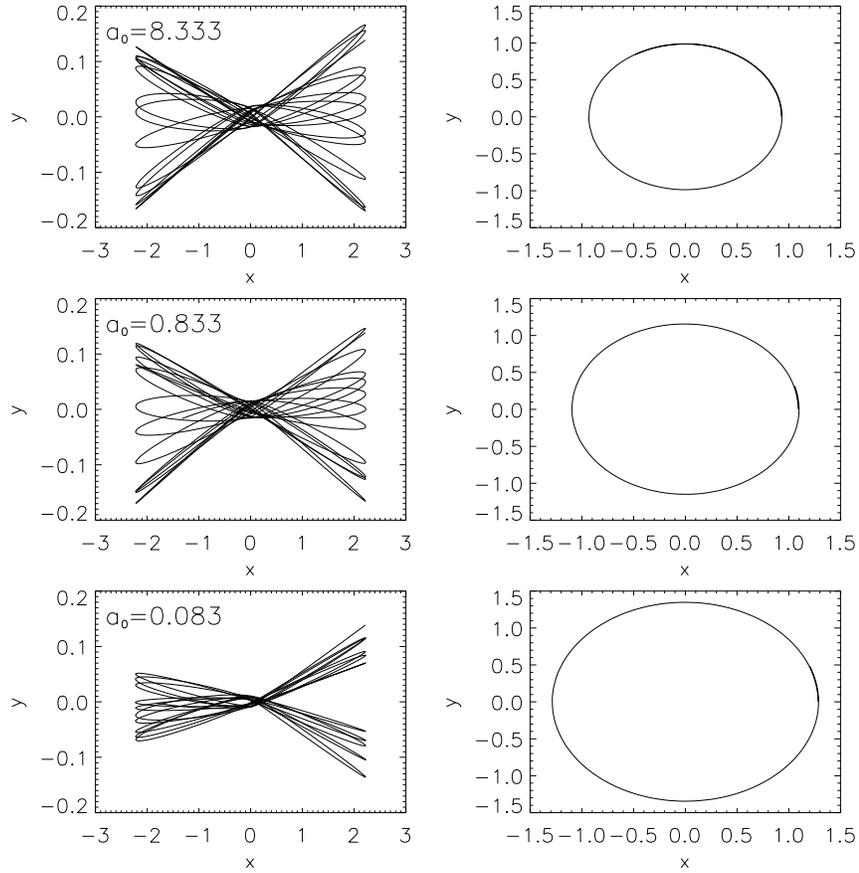} \caption{Orbits in a nonrotating triaxial potential
in MOND. The left panels are for a stationary start space, while the
right panels are for a x-z start space (1:1 resonant orbit). The
energy is from the 10th cell.}\label{orbit}
\end{figure}

Figure \ref{orbit} shows the orbits in three separate planes for
MOND models. The energies of orbits are from shell 10. The orbits in a
triaxial MOND potential are similar to the orbits in a triaixal
Newtonian potential, which yield the box, tube and stochastic
orbits.

\section{Construction of self-consistent models}

Since Schwarzschild (1979) proposed a numerical method to construct
self-consistent models of galaxies, orbit superposition techniques
have become an important tool in dynamical modelling (e.g., Rix
1997; van der Marel et al. 1998; Kuijken 2004; Binney 2005; van de
Ven 2006; Capuzzo-Dolcetta et al. 2006). The procedure of constructing such models can be described as follows:

(1)A three-dimensional mass distribution is employed and the
corresponding gravitational potential is calculated by solving
Poisson's equation. In this paper, the case is revised as we solve
the modified Poisson's equation.  The simulations are also not scale-free,
which was possible for some Newtonian galaxy models.

(2)A full orbital library is established. In other words,
it is necessary to generate a large set of initial conditions for
the orbits.

(3)The whole system is divided into many cells with the same mass.
The fraction of time spent by each orbit in each cell $O_{ij}$ is
recorded into a large matrix.

(4)The non-negative occupation numbers $W_j$ are programmed
in a linear equation group with the above matrix and the desired mass in each cell.
The weights are then solved by non-negative
least squares inversion.
Note that the relation between orbital weights and the density distribution is linear independent of gravity theories.

The linear equation group can be written as

\begin{equation}\label{eq7}
\sum_{j=1}^{N_o}W_jO_{ij}=M_i,~~~~~~~~ i=1,.....,N_c,
\end{equation}
where $M_i$ is the mass of each grid cell, $N_o$ is the number of
orbits, here we have $N_o=6840$. $N_c$ is the number of cells, and
we adopt $N_c$=912. $W_j\geq0$ is the weight number of the
occupation. We follow MF96 to divide the first octant into 960
cells, which can be described by the following steps. The galaxy was
divided by 20 ellipsoidal shells into 21 sections of equal mass. For
each shell, only the first octant was considered. The first octant
was then divided into three parts by the planes $z=h_zx/h_x$,
$y=h_yx/h_x$, and $z=h_zy/h_y$. Each part, for example the first
part, which contained the x-axis, was divided by $h_xy/h_yx=1/5,
2/5, 2/3$ and $h_xz/h_zx=1/5, 2/5, 2/3$ into 16 cells. The total
number of cells is $20\times3\times16=960$. We consider the inner
912 cells in solving the linear equation groups.

Generally, equation (\ref{eq7}) can be solved by linear programming
(e.g., Schwarzschild 1979, 1982, 1993) using Lucy's method (Lucy 1974;
 Statler 1987), or maximum entropy methods (e.g., Richstone \&
Tremaine 1988; Statler 1991; Gebhardt et al. 2003), or with the
least squares solver (Lawson \& Hanson 1974)(MF96; van de Ven et al.
2006; Capuzzo-Dolcetta et al. 2006). The four methods have their own
advantages and disadvantages, which we do not discuss here. We adopt
the least squares method to solve the linear programming. Then the
minimum square deviation can be written as

\begin{equation}\label{eq8}
\chi^2=\frac{1}{N_c}\sum_{j=1}^{N_c}\bigg(C_i-\sum_{i=1}^{N_o}w_jO_{ij}\bigg)^2
\end{equation}

For the symmetry of the triaxial model, only one octant $O_{ij}$ is
considered in solving the linear programming.


\section{Conclusion and discussion}
\begin{figure}
\plotone{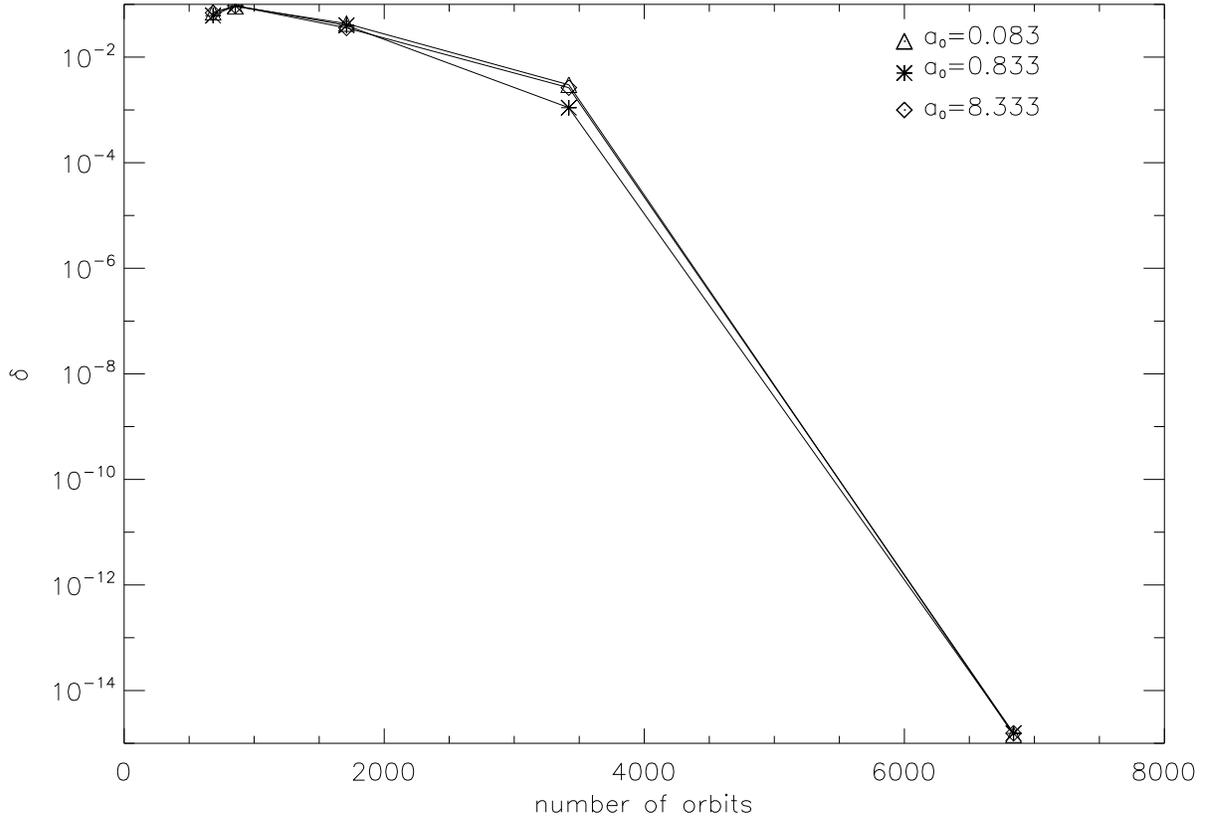} \caption{Departure form self-consistency $\delta$
as a function of the number of orbits for different values of $a_0$. The triangles, asterisks, and diamonds represent $a_0=0.083$, 0.833, and
8.333, respectively.} \label{delta}
\end{figure}

In practice, a parameter $\delta$ is adopted to describe the
departure form self-consistency, which is defined as in MF96:
\begin{equation}
\delta=\frac{\sqrt{\chi^2}}{\bar{M}},
\end{equation}
where $\bar{M}$ is the average mass in each cell. Figure \ref{delta}
shows the departure from self-consistency as a function of the
number of orbits. Note that there is a remarkably strong dependence
on the number of orbits. The departure parameter $\delta$ drops
quickly when the number of orbits exceeds 4000. $\delta$ is
$~10^{-15}$ when the 6840 orbits are adopted in the optimization
routine, which shows that all three triaxial models are
self-consistent in MOND. Amazingly, we also find that the departure
parameter $\delta$ is insensitive to $a_0$. In other words, if one
of the triaxial galaxies is self-consistent in MOND, then the same
model with a different value of $a_0$ will also be self-consistent.

\begin{figure}
\plotone{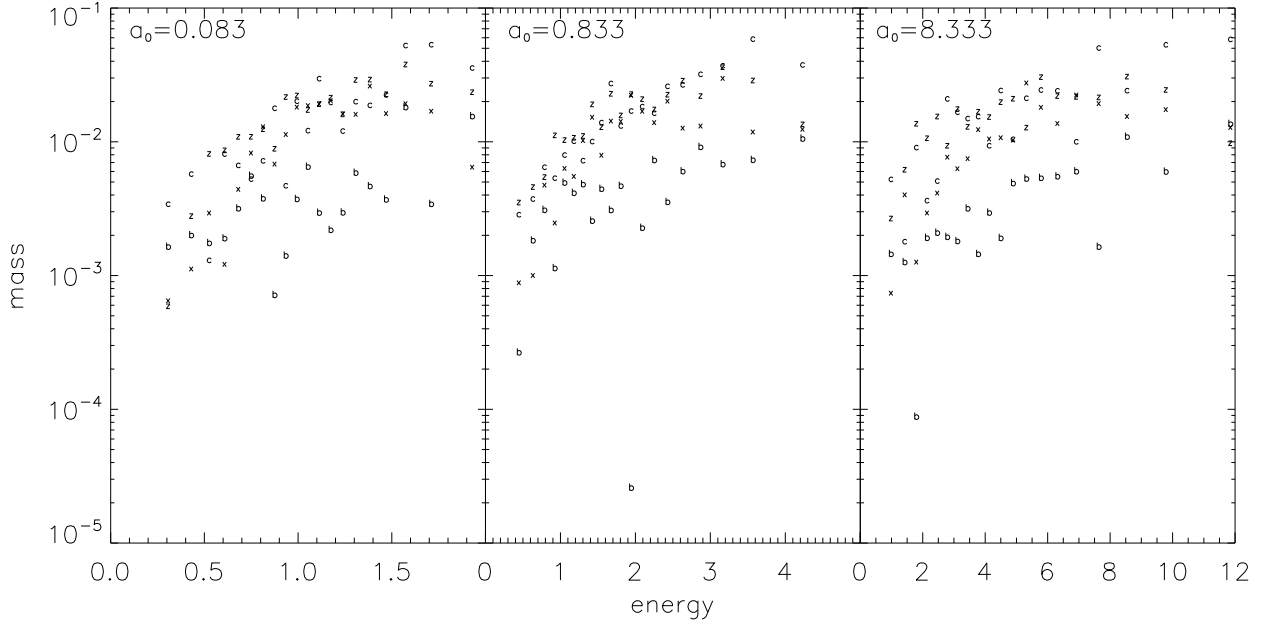} \caption{Orbital families in the self-consistent
models. The symbols b, x, z, and c denote the mass contributed by
box, x-tube, z-tube, and chaotic orbits,
respectively.}\label{ps_den}
\end{figure}

Figure \ref{ps_den} shows the contribution of mass for various
orbital families in the self-consistent model.
Chaotic orbits contribute a larger fraction of mass with the increase of
energy. This shows a certain degree of consistency with that of
Capuzzo-Dolcetta et al. (2007). However, discrepancies between
our results and that of Capuzzo-Dolcetta et al. (2007) are apparent.
There is a clear variational trend of the cumulative energy
distribution of the various orbital families in Newton's case,
however, no apparent variational trend for different orbits has been
found in MOND.

\begin{figure}
\plotone{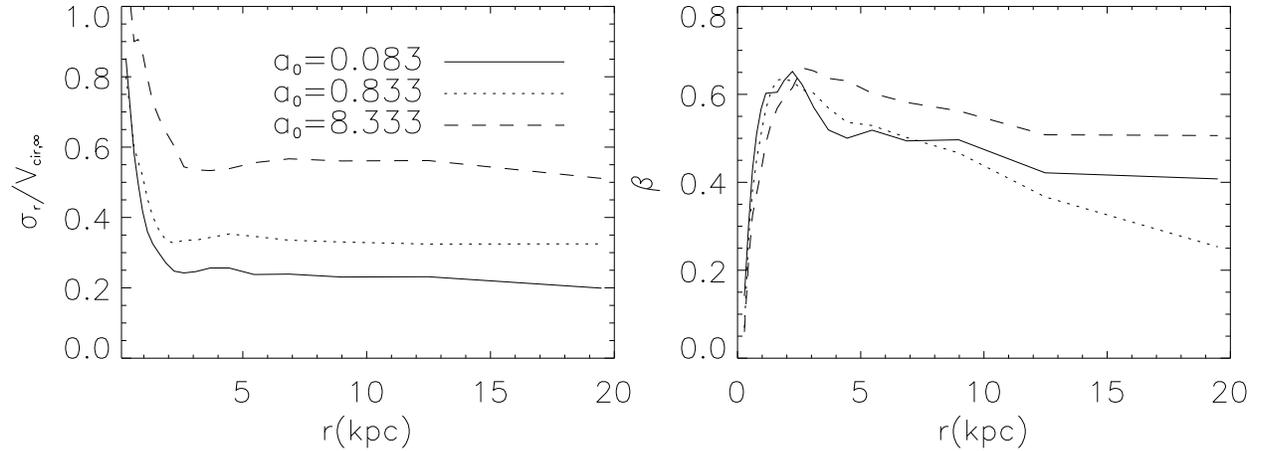} \caption{The radial profiles of the rescaled radial
velocity dispersion $\sigma_r/V_{\rm cir,\infty}$ (left panel) and
of the anisotropy parameter
$\beta=1-0.5<V^2-V_r^2>/<V_r^2>$ (right panel). The solid, dotted
and dashed lines represent $a_0=0.083$, 0.833, 8.333,
respectively.}\label{sigma}
\end{figure}

In Figure \ref{sigma}, we show the radial profile of the rescaled
radial velocity dispersion $\sigma_r/V_{\rm cir,\infty}$(left panel)
and the radial profile of the anisotropy parameter
$\beta=1-0.5(<V^2-V_r^2>)/<V_r^2>$ (right panel), where $V_{\rm
cir,\infty}$ is the circular velocity at infinity, and $V_r$ is
radial velocity. The parameter $V^2$ is defined as
$V^2=v_x^2+v_y^2+v_z^2$, where $v_x$, $v_y$ and $v_z$ are the three
components of velocity. It can be seen that the profiles of the
radial velocity dispersion are nearly flat in MOND, except in the
central region. Recently, Capuzzo-Dolcetta et al. (2007) studied the
same model as that used here, but in the Newtonian gravity. It can
be seen that the profile of the velocity dispersion in our model
shows a certain degree of consistency with that of Capuzzo-Dolcetta
et al. (2007), but the discrepancies between our results and that of
Capuzzo-Dolcetta et al. (2007) are apparent. There is central peak
in our velocity dispersion profile. The global average of the
velocity  $V_{\rm rms}$ is 170.49, 78.84 and 42.70 km/s for
$a_0$=0.083, 0.833, and 8.333, respectively (see the last column of
Table~\ref{parameter}).

The anisotropy parameter $\beta\geq0$ suggests that the stars have
high radial velocity and the ratio of $<V_r^2>/<V^2-V_r^2>$ reaches
a maximum value at $2-3$ kpc. Our anisotropy profiles are similar to
that of Nipoti et al. (2007b) form dissapationless collapse in MOND
(their figur 2) and in Newtonian gravity van Albada (1982) and Wang
et al. (2008a). A gently rising beta is widely used (e.g., Milgrom
\& Sanders 2003). However, the peak in the anisotropy profile in our
model is not seen in any simulations and observations to our best
knowledge. This is might be indication of problem of MOND. Our
models have a sudden change of the amount of box orbit near shells
10-12 (radius~2-3kpc), which could be the reason.

\begin{figure}
\plotone{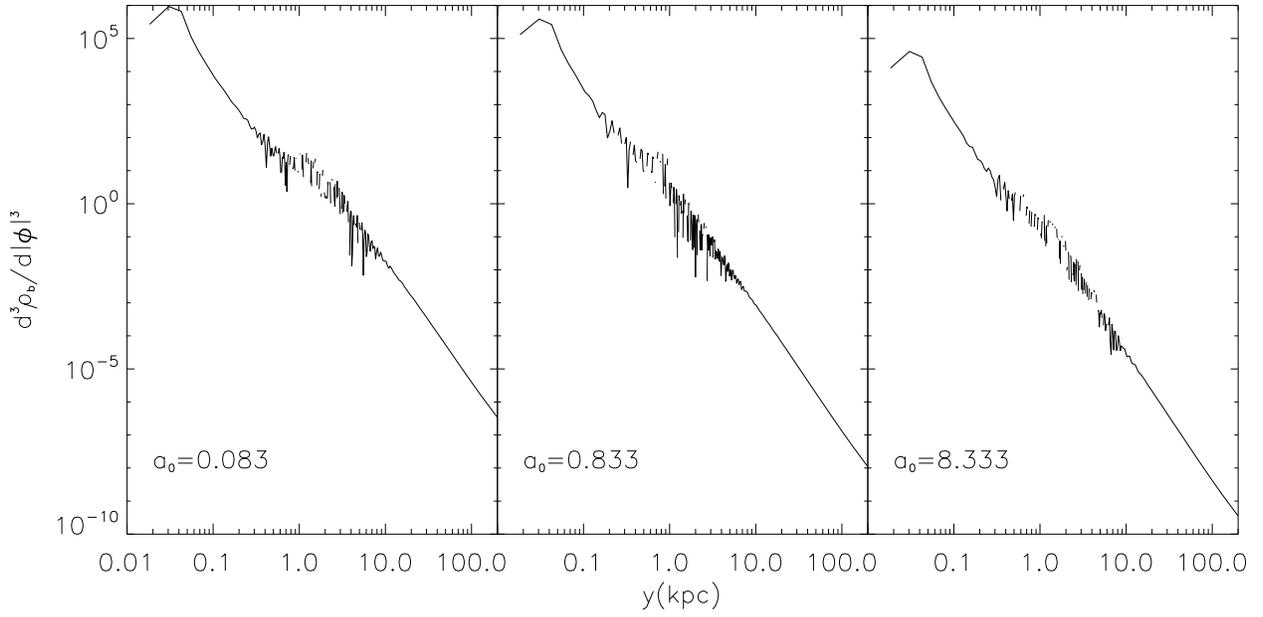} \caption{Distribution of ${\rm
d}^3\rho_b(0,y,0)/{\rm d}|\Phi(0,y,0)| ^3 $ as function of MOND
potential $|\Phi(0,y,0|)$ along the intermediate axis for our three
models.}\label{drdp}
\end{figure}

In order to check the stability of our models, we used the Antonov's
third law (Binney \& Tremaine 1987). If the stellar system is
stable, then its density $\rho$ and potential $\Phi$ satisfy
everywhere the inequality ${\rm d}^3\rho_b/{\rm d}\Phi^3 < 0$.
Figure~\ref{drdp} shows that ${\rm d}^3\rho_b(0,y,0)/{\rm
d}|\Phi(0,y,0)|^3$ is positive everywhere along the intermediate
axis for our three models, since the values of potential are
negative, our models are consistent with the law, which means the
models appear globally stable. We will also be checking the
stability of the models in collaboration with groups which have MOND
N-body code. Effects due to pattern rotation and external field in
MOND remain to be studied (Wang et al. 2008b).

We have used the Schwarzschild approach to examine whether the triaxial
galaxies are self-consistent in MOND. Using the Bologna Poisson
solver to determine the potential and the accelerations at some discrete
points, we used three-dimensional interpolation then obtain the
potential and accelerations at arbitrary point. The orbits conserve
energy to a very high level of accuracy, which shows that our method
is feasible in calculating orbits using a grid-based MONDian potential.
The departure parameter $\delta$ shows that the triaxial galaxy
models adopted in this paper in MOND are self-consistent. Our
results show that it is theoretically allowed to have triaxial galaxies
in rigorous equilibrium in a universe with either
non-Newtonian gravity or certain models of Dark Energy (Zhao 2007).

\acknowledgements We thank the referee for the constructive comments
which improved our presentation, Luca Ciotti, Pasquale Londrio,
Carlo Nipoti for generously sharing their MONDian Poisson solver
code and Dr. Scott Gregory for assistance on written English, Zuhui
Fan for discussions. YGW is supported by the National Science
Foundation of China under grants 10525314, 10533010, by the Chinese
Academy of Sciences under grant KJCX3-SYW-N2, and by the Ministry of
Science and Technology under the national basic sciences program
(973) under grant 2007CB815401. HZ acknowledges partial support of
PPARC Advanced Fellowship and National Natural Science Foundation of
China (NSFC under grant 10428308). XW acknowledges the support of
SUPA studentship.

\begin{table}
 \caption {Main physical parameters for our three ellipsoidal MOND models}\label{parameter}
\begin{center}
\begin{tabular}{llllllllll}\hline\hline
Model&$a_0^a$[km/s/Gyr]&$M[M_{\odot}]$&$V_{\rm
cir,\infty}$[km/s]&$V_{\rm rms}$[km/s]\\
\hline
$a_0=0.083$&3600&$10^{11}$&198.6&170.5\\
$a_0=0.833$&3600&$10^{10}$&111.7&78.8\\
$a_0=8.333$&3600&$10^9$   &62.8 &42.7\\
\hline
\end{tabular}
\end{center}
{\footnotesize
 \noindent
 $^{a}$ real value of $a_0$.\\}

\end{table}

\begin{table}
 \caption {The energy and dynamical time in every shell}\label{shell}
\begin{center}
\begin{tabular}{llllllllll}\hline\hline
&Radius$^{a}$&&Energy&&&$T_D$\\
\cline{2-2}\cline{3-5}\cline{6-8}
Shell&$a_0$$^{b}$&$a_0=0.083$&$a_0=0.833$&$a_0=8.333$&$a_0=0.083$&$a_0=0.833$&$a_0=8.333$\\
\hline
1&0.2791&0.2933&0.4113&0.9006&2.640&2.200&1.477\\
2&0.4464&0.4163&0.5990&1.341&3.600&2.955&1.960\\
3&0.6076&0.5120&0.7536&1.713&4.485&3.645&2.395\\
4&0.7744&0.5940&0.8908&2.055&5.400&4.335&2.815\\
5&0.9529&0.6675&1.021&2.384&6.400&5.065&3.255\\
6&1.148&0.7355&1.145&2.706&7.470&5.815&3.720\\
7&1.366&0.7995&1.267&3.030&8.685&6.705&4.255\\
8&1.612&0.8610&1.390&3.358&10.16&7.530&4.810\\
9&1.896&0.9209&1.513&3.695&11.71&8.770&5.490\\
10&2.226&0.9800&1.639&4.046&13.71&10.05&6.220\\
11&2.620&1.039&1.771&4.416&16.04&11.56&7.120\\
12&3.097&1.099&1.910&4.810&19.00&13.47&8.185\\
13&3.690&1.160&2.057&5.237&22.72&15.80&9.510\\
14&4.449&1.225&2.217&5.705&27.68&18.76&11.23\\
15&5.458&1.294&2.395&6.231&34.23&22.71&13.64\\
16&6.866&1.370&2.597&6.834&44.00&28.59&16.76\\
17&8.974&1.456&2.835&7.554&58.23&39.75&21.31\\
18&12.48&1.560&3.131&8.459&82.40&50.47&29.12\\
19&19.49&1.697&3.533&9.702&131.2&78.07&44.72\\
20&40.49&1.915&4.196&11.77&279.0&161.6&91.36\\
 \hline
 \end{tabular}
\end{center}
 {\footnotesize
 \noindent
 $^{a}$ along the x-axis.\\}
 $^{b}$ $a_0=0.083, 0.833,8.333$.\\
\end{table}

\end{document}